%% file: 2l5pt-lett.tex
\documentclass[aps,prl,10pt,nofootinbib,reprint,superscriptaddress,preprintnumbers,floatfix]{revtex4-2}
\usepackage[nodisplayskipstretch]{setspace}
\usepackage{mathtools}
\usepackage{booktabs}
\usepackage{xcolor}
\usepackage{amssymb}
\usepackage{amsmath}
\usepackage{mathrsfs}
\usepackage{microtype}
\usepackage[hidelinks]{hyperref}
\definecolor{desycyan}{rgb}{0.00,0.68,0.93}

\usepackage{tikz}
\usetikzlibrary{arrows.meta}
\usetikzlibrary{decorations.pathmorphing}

\usepackage[english]{babel}
\makeatletter
\newcommand{\letterappendixtitle}{%
  \par
  \onecolumngrid@push
  \addvspace{0.8cm plus 1ex minus .2ex}%
  \noindent
  \hbox to\textwidth{%
    \hfil\normalfont\large\bfseries Appendix\hfil
  }%
  \par
  \addvspace{0.5cm}%
  \onecolumngrid@pop
}
\makeatother

\definecolor{desyorange}{rgb}{0.96,0.52,0.07}
\definecolor{internationalorange}{rgb}{1.0,0.31,0.0}

\newcommand{\dd}{\mathrm{d}}
\newcommand{\ii}{\mathrm{i}}
\newcommand{\Li}{\operatorname{Li}_2}
\newcommand{\PV}{\operatorname{PV}}

\tikzset{
  matter/.style={thick},
  photon/.style={draw=desyorange,decorate,decoration={snake,amplitude=.5mm,segment length=2.2mm}},
  radiation/.style={photon,draw=internationalorange},
  cut/.style={draw=black,dash pattern=on 1.6pt off 1.2pt,line width=.5pt},
}

\begin{document}

\preprint{USTC-ICTS/PCFT-26-87}

\title{\sffamily\bfseries Analytic Two-Loop Classical Bremsstrahlung Amplitude in Electrodynamics}

\author{Xinguang Chen}
\affiliation{Department of Physics and Astronomy, Johns Hopkins University, Baltimore, MD 21218, USA}

\author{Zhengwen Liu}
\affiliation{School of Physics and Shing-Tung Yau Center, Southeast University, Nanjing 210096, China}

\author{Rourou Ma}
\affiliation{Interdisciplinary Center for Theoretical Study, University of Science and Technology of China, Hefei 230026, China}
\affiliation{Max-Planck-Institut f\"ur Physik, Werner-Heisenberg-Institut, Boltzmannstraße 8, 85748 Garching, Germany}

\author{Yu Wu}
\affiliation{LAPTh, Université Savoie Mont Blanc, CNRS, B.P. 110, F-74941 Annecy-le-Vieux, France}

\author{Yang Zhang}
\affiliation{Interdisciplinary Center for Theoretical Study, University of Science and Technology of China, Hefei 230026, China}
\affiliation{Peng Huanwu Center for Fundamental Theory, Hefei 230026, China}
\affiliation{Center for High Energy Physics, Peking University, Beijing 100871, China}

\begin{abstract}
We compute the leading long-wavelength term of the two-loop five-point amplitude for photon emission in the scattering of two massive spinless charges. Classical consistency suggests that two Coulomb exchanges dress the radiative tree amplitude. We show that this iteration accounts for the entire leading term and evaluate the resulting convolution. At fixed external transfer, the exchanges shift the momentum in the tree numerator, thereby changing the relative weights of its two polarization structures. We determine both coefficients at nonzero photon frequency and general relative velocity, with exact dependence on the spacetime dimension.
\end{abstract}
\maketitle

\noindent{\bf Introduction.}~Gravitational-wave observations have established merging black holes as a testing ground for general relativity in its strong-field, dynamical regime~\cite{LIGOScientific:2016aoc,LIGOScientific:2026qni}.
Inferring masses and spins from these signals requires accurate models of the binary's motion and radiation~\cite{Peters:1964zz,Purrer:2019jcp}. During a quasicircular inspiral, the frequency evolution is governed by the radiated power and the change in binding energy per unit frequency~\cite{Blanchet:2023bwj}.
At the high signal-to-noise ratios attainable with next-generation detectors, even small waveform errors can bias these estimates by more than their statistical uncertainties~\cite{Purrer:2019jcp}.

\vskip 2pt
Quantum field theory offers a systematic route from relativistic scattering to binary dynamics. The post-Minkowskian (PM) expansion keeps the relative velocity fixed at each order in Newton's constant. For conservative motion, analytic continuation relates the scattering angle to the periastron advance of a bound orbit~\cite{Kalin:2019rwq,Kalin:2019inp}.
Combining effective field theory~\cite{Goldberger:2004jt,Kalin:2020mvi} with $S$-matrix techniques~\cite{Bjerrum-Bohr:2018xdl,Cheung:2018wkq,Kosower:2018adc}, several groups have determined the conservative dynamics of nonspinning bodies at 3PM~\cite{Bern:2019nnu,Bern:2019crd,Kalin:2020fhe} and 4PM~\cite{Bern:2021dqo,Dlapa:2021npj,Bern:2021yeh,Dlapa:2021vgp,Dlapa:2024cje}.
The 5PM potential contribution has been extracted from four-loop elastic scattering for general mass ratios~\cite{Bern:2025wyd}. The conservative tail contribution to the deflection, due to radiation scattered by the binary's gravitational field, has also been computed in a mass-ratio expansion~\cite{Driesse:2024xad,Dlapa:2025biy,Dlapa:2026oyq}. For quasicircular, nonspinning inspirals, complementary post-Newtonian (PN) calculations yield the flux through 4PN and the phase through 4.5PN~\cite{Blanchet:2023sbv,Blanchet:2023bwj}.

\vskip 2pt
In scattering, momentum conservation relates the total radiated four-momentum to the two bodies' impulses, as in the 4PM calculation of radiation reaction~\cite{Dlapa:2022lmu,Dlapa:2023hsl}. To resolve the frequency, direction, and polarization in an amplitude calculation, we keep the emitted graviton as a fifth external particle alongside the two incoming and two outgoing bodies.
Fourier transforming the tree amplitude in the momentum transfer gives the leading scattering waveform~\cite{Kovacs:1978eu,Jakobsen:2021smu,Cristofoli:2021vyo}. The next correction combines the one-loop amplitude with products of tree amplitudes~\cite{Herderschee:2023fxh,Elkhidir:2023dco,Caron-Huot:2023vxl} and has been evaluated analytically at general relative velocity~\cite{Brunello:2025eso}.
For nonspinning bodies with a general mass ratio, the two-loop waveform is known through relative 2PN order~\cite{Bini:2024ijq}, and its even-parity quadrupole and octupole through 3.5PN~\cite{Bini:2026dvn,Bini:2026jwc}. Since even small deflections can occur at relativistic speeds, a key input for extending this prediction to general velocities is the two-loop five-point amplitude.

\vskip 2pt
In electrodynamics the corresponding five-point amplitude describes bremsstrahlung from two scattering charges~\cite{Cristofoli:2021vyo,Elkhidir:2023dco}. The spectrum and polarization are sensitive to how the charges accelerate throughout the encounter~\cite{Jackson:1998nia,Saketh:2021sri}.
A charge can exchange momentum with its partner both before and after emitting a photon. Summing these orderings can put intermediate matter lines on shell in the long-wavelength expansion. Although graviton self-interactions generate additional diagrams, the same identities apply to massive lines in gravity~\cite{Brandhuber:2023hhy}. In electrodynamics, classical consistency relates one-loop emission to elastic and radiative tree amplitudes and suggests an iteration with two Coulomb exchanges at two loops~\cite{Cristofoli:2021jas}.

\vskip 2pt
In this work, we compute the classical two-loop five-point bremsstrahlung amplitude for two massive spinless charges at leading long-wavelength order. When expressed in terms of the physical masses and charges, the full diagram sum reduces to a convolution of the leading radiative tree amplitude with two Coulomb exchanges. Each exchange shifts the momentum transfer in both the tree propagator and its polarization numerator. We evaluate the convolution at finite photon frequency and general relative velocity, obtaining both polarization coefficients with exact dependence on the spacetime dimension.
For opposite charges, we exhibit an exact tree zero where the double infrared pole vanishes, but the convolution samples other transfers and the single-pole residues from both emitting lines reinforce each other.

\newpage
\vskip 6pt
\noindent{\bf Long-wavelength expansion.}~We study bremsstrahlung in scalar electrodynamics with two complex fields of masses $m_a$ and charges $Q_a$ (including the coupling). The Lagrangian is
\begin{align}\label{eq:lagrangian}
\mathcal L={}&-\frac14F_{\mu\nu}F^{\mu\nu}\nonumber\\
&+\sum_{a=1}^2\left[\big|(\partial_\mu+\ii Q_aA_\mu)\phi_a\big|^2-m_a^2|\phi_a|^2\right].
\end{align}
In $D=4-2\epsilon$ the charges are $\mu^\epsilon Q_a$, with $\mu$ denoting the scale of dimensional regularization. We strip off the tree-level factors $\mu^{3\epsilon}$ for emission and $\mu^{2\epsilon}$ for elastic scattering, respectively. In the loop integrals we suppress $\mu^{2\epsilon}$ per loop and restore it when extracting Laurent coefficients. We use Feynman gauge, the mostly-minus metric and an implicit $-\ii0$ for uncut propagators.

\vskip 4pt
In the long-wavelength regime the exchanged and emitted momenta are small compared with the particle masses, even at relativistic speeds. We describe each massive line by its average momentum $M_au_a$ and transfer $q_a$,
\begin{equation}\label{eq:kinematics}
\begin{aligned}
\begin{tikzpicture}[baseline={(0,0)},scale=.58,
  massive/.style={line width=.85pt},
  mom/.style={line width=.5pt,-{Latex[length=3pt,width=2pt]}}]
\filldraw[fill=gray!45,draw=black,line width=.5pt] (0,0) ellipse [x radius=.42,y radius=.75];
\draw[massive] (-1.55,-1.2)--(-.30,-.525);
\draw[massive] (-1.55,1.2)--(-.30,.525);
\draw[massive] (.30,.525)--(1.55,1.2);
\draw[massive] (.30,-.525)--(1.55,-1.2);
\draw[radiation] (.42,0)--(1.9,0);
\draw[mom] (-1.094,-1.159)--(-.610,-.897);
\draw[mom] (-1.094,1.159)--(-.610,.897);
\draw[mom] (.610,.897)--(1.094,1.159);
\draw[mom] (.610,-.897)--(1.094,-1.159);
\node[font=\small,inner sep=1pt,anchor=north east] at (-1.64,-1.25) {$p_2$};
\node[font=\small,inner sep=1pt,anchor=south east] at (-1.64,1.25) {$p_1$};
\node[font=\small,inner sep=1pt,anchor=south west] at (1.64,1.25) {$p_3$};
\node[font=\small,inner sep=1pt,anchor=north west] at (1.64,-1.25) {$p_4$};
\node[font=\small] at (2.18,0) {$k$};
\end{tikzpicture}
\end{aligned}
\quad
\begin{aligned}
p_1&=M_1u_1+\tfrac12q_1,\\
p_3&=M_1u_1-\tfrac12q_1,\\
p_2&=M_2u_2+\tfrac12q_2,\\
p_4&=M_2u_2-\tfrac12q_2.
\end{aligned}
\end{equation}
Momentum conservation implies $q_1+q_2=k$. Each transfer is spacelike, with $t_a=-q_a^2$.
With $u_a^2=1$, the mass-shell conditions imply $u_a\cdot q_a=0$ and $M_a^2=m_a^2+t_a/4$ for a pole mass $m_a$.
The relative speed is $\beta/y$, where $y=u_1\cdot u_2>1$ and $\beta=\sqrt{y^2-1}$. The photon energy in the rest frame of $u_a$ is $\omega_a=u_a\cdot k>0$. We write $\Delta=t_1-t_2$, $E_a=\varepsilon\cdot u_a$ and $E_q=\varepsilon\cdot q_2$.

\vskip 4pt
In the soft region we scale $q_a,k,\ell_b\mapsto\lambda(q_a,k,\ell_b)$ together at fixed $m_a,u_a,\varepsilon$~\cite{Beneke:1997zp,Parra-Martinez:2020dzs}. For soft $\ell$, the inverse scalar propagator $m_a^2-(p_{a,\mathrm{in}}+\ell)^2=-2M_au_a\cdot\ell-\ell\cdot(\ell+q_a)$ has a linear leading term. Expanding about this term yields the eikonal propagator and its first correction.

\vskip 4pt
Individual one- and two-loop integrands can be as singular as $\lambda^{-7}$ and $\lambda^{-11}$, respectively. At positive photon energy these terms cancel in the sum or integrate to zero. We denote the leading surviving term of the amplitude by $\mathcal M_5^{(L)}$. Its integrand degree is $-2,-6,-10$ at tree level, one loop, and two loops, respectively. Including the loop measures gives a degree of $-2-2L\epsilon$. With soft momenta proportional to $\hbar$ and $Q_a\propto\hbar^{-1/2}$, the amplitude scales as $\mathcal M_5^{(L)}\propto\hbar^{-7/2-L}$ in four dimensions~\cite{Kosower:2018adc,Cristofoli:2021jas}.
The two-loop term computed here is therefore enhanced by $\hbar^{-2}$ relative to the radiative tree amplitude.

\vskip 4pt
At tree level, the two orderings of emission and exchange cancel at order $\lambda^{-3}$ on each line. At order $\lambda^{-2}$ the two-photon contact (seagull) vertex also contributes. The sum can be written in terms of two contractions of the emitted photon's field strength,
\begin{align}\label{eq:WV}
\mathcal W&=\Delta E_1-2\omega_1E_q=2\,q_2^\mu F_{\mu\nu}u_1^\nu,\nonumber\\
\mathcal V&=\omega_2E_1-\omega_1E_2=u_2^\mu F_{\mu\nu}u_1^\nu,
\end{align}
where $F_{\mu\nu}=k_\mu\varepsilon_\nu-k_\nu\varepsilon_\mu$ and momentum conservation gives $\Delta=2k\cdot q_2$.
At fixed $k,u_1,u_2$, only $\mathcal W$ depends on the momentum transfer. The leading tree amplitude is
\begin{equation}\label{eq:tree}
\mathcal M^{(0)}_5=\frac{2Q_1^2Q_2m_2}{t_2}\left[\frac{y\mathcal W}{\omega_1^2}-\frac{2\mathcal V}{\omega_1}\right]+(1\leftrightarrow2).
\end{equation}
Throughout, $(1\leftrightarrow2)$ exchanges all particle labels, with $E_q\to-E_q$ following from $\varepsilon\cdot k=0$.

\begin{figure}[t]
\centering
\begin{tikzpicture}[every node/.style={font=\small},
  mline/.style={line width=.75pt},
  sphoton/.style={draw=desyorange,line width=.45pt,decorate,decoration={snake,amplitude=.3mm,segment length=1.4mm}},
  sradiation/.style={sphoton,draw=internationalorange}]
\begin{scope}[shift={(0.00,0)}]
\draw[sphoton,decoration={segment length=0.131cm}] (0.1500,0.9200)--(0.1500,0.0000);
\draw[sphoton,decoration={segment length=0.131cm}] (0.6700,0.9200)--(0.6700,0.0000);
\draw[sphoton,decoration={segment length=0.131cm}] (0.4100,0.9200)--(0.4100,0.0000);
\draw[sradiation] (.41,.92)--(.61,1.19);
\draw[mline] (0,.92)--(.82,.92);
\draw[mline] (0,0)--(.82,0);
\end{scope}
\begin{scope}[shift={(1.16,0)}]
\draw[sphoton,decoration={segment length=0.131cm}] (0.1500,0.9200)--(0.1500,0.0000);
\draw[sphoton,decoration={segment length=0.121cm}] (0.6700,0.9200)--(0.5712,0.5703);
\draw[sphoton,decoration={segment length=0.121cm}] (0.5088,0.3497)--(0.4100,0.0000);
\draw[sphoton,decoration={segment length=0.137cm}] (0.4100,0.9200)--(0.6700,0.0000);
\draw[sradiation] (.41,.92)--(.61,1.19);
\draw[mline] (0,.92)--(.82,.92);
\draw[mline] (0,0)--(.82,0);
\end{scope}
\begin{scope}[shift={(2.32,0)}]
\draw[sphoton,decoration={segment length=0.121cm}] (0.1500,0.9200)--(0.2488,0.5703);
\draw[sphoton,decoration={segment length=0.121cm}] (0.3112,0.3497)--(0.4100,0.0000);
\draw[sphoton,decoration={segment length=0.131cm}] (0.6700,0.9200)--(0.6700,0.0000);
\draw[sphoton,decoration={segment length=0.137cm}] (0.4100,0.9200)--(0.1500,0.0000);
\draw[sradiation] (.41,.92)--(.61,1.19);
\draw[mline] (0,.92)--(.82,.92);
\draw[mline] (0,0)--(.82,0);
\end{scope}
\begin{scope}[shift={(3.48,0)}]
\draw[sphoton,decoration={segment length=0.138cm}] (0.1500,0.9200)--(0.3004,0.3880);
\draw[sphoton,decoration={segment length=0.117cm}] (0.3463,0.2254)--(0.4100,0.0000);
\draw[sphoton,decoration={segment length=0.134cm}] (0.6700,0.9200)--(0.5382,0.6869);
\draw[sphoton,decoration={segment length=0.155cm}] (0.4551,0.5398)--(0.1500,0.0000);
\draw[sphoton,decoration={segment length=0.137cm}] (0.4100,0.9200)--(0.6700,0.0000);
\draw[sradiation] (.41,.92)--(.61,1.19);
\draw[mline] (0,.92)--(.82,.92);
\draw[mline] (0,0)--(.82,0);
\end{scope}
\begin{scope}[shift={(4.64,0)}]
\draw[sphoton,decoration={segment length=0.134cm}] (0.1500,0.9200)--(0.2818,0.6869);
\draw[sphoton,decoration={segment length=0.183cm}] (0.3649,0.5398)--(0.4551,0.3802);
\draw[sphoton,decoration={segment length=0.134cm}] (0.5382,0.2331)--(0.6700,0.0000);
\draw[sphoton,decoration={segment length=0.137cm}] (0.6700,0.9200)--(0.4100,0.0000);
\draw[sphoton,decoration={segment length=0.137cm}] (0.4100,0.9200)--(0.1500,0.0000);
\draw[sradiation] (.41,.92)--(.61,1.19);
\draw[mline] (0,.92)--(.82,.92);
\draw[mline] (0,0)--(.82,0);
\end{scope}
\begin{scope}[shift={(5.80,0)}]
\draw[sphoton,decoration={segment length=0.135cm}] (0.1500,0.9200)--(0.3500,0.5662);
\draw[sphoton,decoration={segment length=0.135cm}] (0.4700,0.3538)--(0.6700,0.0000);
\draw[sphoton,decoration={segment length=0.135cm}] (0.6700,0.9200)--(0.4700,0.5662);
\draw[sphoton,decoration={segment length=0.135cm}] (0.3500,0.3538)--(0.1500,0.0000);
\draw[sphoton,decoration={segment length=0.131cm}] (0.4100,0.9200)--(0.4100,0.0000);
\draw[sradiation] (.41,.92)--(.61,1.19);
\draw[mline] (0,.92)--(.82,.92);
\draw[mline] (0,0)--(.82,0);
\end{scope}
\foreach \x in {.99,2.15,3.31,4.47,5.63} {\node at (\x,.46) {$+$};}
\draw[-{Stealth[length=4.6pt,width=4.2pt]},line width=1.1pt,draw=black!70] (6.78,.46)--(7.30,.46);
\begin{scope}[shift={(7.46,0)}]
\draw[sphoton,decoration={segment length=0.131cm}] (0.1500,0.9200)--(0.1500,0.0000);
\draw[sphoton,decoration={segment length=0.131cm}] (0.4100,0.9200)--(0.4100,0.0000);
\draw[sphoton,decoration={segment length=0.131cm}] (0.6700,0.9200)--(0.6700,0.0000);
\draw[sradiation] (.41,.92)--(.61,1.19);
\draw[mline] (0,.92)--(.82,.92);
\draw[mline] (0,0)--(.82,0);
\foreach \c in {.28,.54} {\draw[cut] (\c,-.127)--(\c,.127);}
\end{scope}
\end{tikzpicture}
\vskip -1mm
\caption{A representative ordering sum on line 2. The upper line emits the photon $k$ at a contact vertex, and the three exchanged photons carry momenta $h_j$ with $\sum_j h_j=q_2$, so $\xi_j=u_2\cdot h_j$ satisfy $\sum_j\xi_j=0$. For a fixed order on line 1, summing the six orderings on line 2 gives $(2\pi\ii)^2\delta(\xi_1)\delta(\xi_2)$ at eikonal order, up to common factors. The dashed lines on the right mark the two cuts, which define sector $22$.}
\label{fig:cuts}
\vskip -10pt
\end{figure}
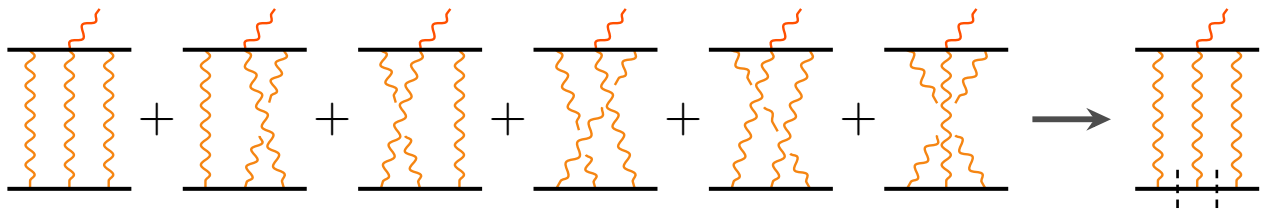

\vskip 6pt
\noindent{\bf From diagrams to transverse exchange.}~We integrate over the full soft region at fixed relative velocity. The transverse constraints on the two additional exchanges follow from summing their attachment orders on the massive lines. Summing two opposite orderings puts the intermediate massive line on its linearized mass shell through the identity
\begin{equation}\label{eq:eikonal}
\frac{1}{\xi-\ii0}+\frac{1}{-\xi-\ii0}=2\pi\ii\,\delta(\xi),
\end{equation}
where $\xi$ is a linearized inverse propagator. The six orderings in Fig.~\ref{fig:cuts} give two such delta functions, or matter cuts~\cite{Levy:1969cr,Abarbanel:1969ek,Cheng:1969eh}. Higher propagator powers lead to derivatives of these cuts.

\vskip 4pt
Of the 3322 two-loop diagrams generated with QGRAF~\cite{Nogueira:1991ex}, 1288 contain closed massive loops. When both loop momenta are soft, these diagrams have integrand degree at least $-7$ because each massive propagator on a closed loop is of order one and each cubic vertex on it supplies a soft momentum. In hard subgraphs, charge conjugation cancels odd-photon currents, the definition of the charge in the Thomson limit absorbs the constant part of the vacuum polarization, and higher local photon interactions carry extra soft momenta~\cite{Grozin:2009an}. Of the remaining 2034 diagrams, 280 have every internal photon connecting the two lines. In 32 of these diagrams, two photons join the same pair of seagulls, giving an integrand degree above $-10$. We sum the other 248 over attachment orders. With physical masses and charges and the external-line normalization specified in the Appendix, diagrams in which a photon returns to the same line give no net contribution at this order. These include the mixed sectors $Q_1^5Q_2^2$ and $Q_1^2Q_2^5$, with one matter cut on each line. After scaleless terms are dropped, cut-preserving integration by parts leaves two terms in each mixed sector with identical denominators, equal coefficients and opposite numerators. Only $11$ and $22$ survive, with both cuts on line 1 or line 2 and charge factors $Q_1^3Q_2^4$ or $Q_1^4Q_2^3$, respectively. Exchanging particle labels interchanges the sectors.

\vskip 4pt
In sector $22$ the two cuts set the loop energies to zero in the rest frame of $u_2$. To combine the remaining linear denominators on the emitting line, we sum its attachment orders with their causal prescriptions and relabel exchanges under the integral. At $\omega_1>0$ this yields $\delta(u_1\cdot h_1)\delta(u_1\cdot h_2)$ (Appendix). The two extra exchanges thus carry no energy in either rest frame and lie in the $d_\perp=D-2$ dimensions transverse to both velocities. The remaining exchange, $h_3=q_2-h_1-h_2$, carries the longitudinal transfer $u_1\cdot h_3=\omega_1$ into the radiative tree, leaving only the three photon denominators.

\vskip 6pt
\noindent{\bf The two-loop amplitude.}~The two extra exchanges change the relative weights of $\mathcal W$ and $\mathcal V$ through a scalar transverse integral $\mathcal T$ and its first moment $\mathcal N$ along the transverse transfer. At fixed $y,t_2$, the frequency dependence of both moments in sector $22$ enters through $z=\omega_1/(\beta\sqrt{t_2})$. In the rest frame of $u_2$, $\omega_1=u_1\cdot q_2=-\boldsymbol u_1\cdot\boldsymbol q_2$, so $z$ is the cosine of the angle between $\boldsymbol u_1$ and $-\boldsymbol q_2$. Positive photon energy restricts $z$ to $0<z\leq1$. For $0<z<1$ we find
\begin{align}\label{eq:amplitude}
\left.\mathcal M^{(2)}_5\right|_{22}={}&\frac{8\pi^2Q_1^4Q_2^3m_2\,y^3}{2^{2D}\pi^D\,\omega_1^2}\bigg[\mathcal W\Big(\mathcal T-\frac{\mathcal N}{t_2(1-z^2)}\Big)\nonumber\\
&-\frac{2\omega_1}{y}\,\mathcal V\Big(\mathcal T-\frac{y^2\mathcal N}{\beta^2t_2(1-z^2)}\Big)\bigg].
\end{align}
The full amplitude $\mathcal M^{(2)}_5$ is obtained by adding the $(1\leftrightarrow2)$ term, with angular variable $\omega_2/(\beta\sqrt{t_1})$. Replacing $\varepsilon$ by $k$ sets $F_{\mu\nu}=0$ in \eqref{eq:WV}, so each charge sector separately satisfies the photon Ward identity.
Before evaluating the master integrals, we also verified that all nine reduced coefficients in \eqref{eq:coefficients} vanish individually under $\varepsilon\to k$.

\vskip 4pt
In \eqref{eq:amplitude}, $\mathcal T$ multiplies the tree combination $\mathcal W-2\omega_1\mathcal V/y$. Only $\mathcal N$ alters the relative weights.

\vskip 4pt
For $0<z<1$ both moments converge absolutely in the strip $-1/4<\operatorname{Re}\epsilon<0$. We evaluate them there and continue analytically in $D$ (Appendix). The exact results at this long-wavelength order are
\begin{align}\label{eq:allEpsilon}
\mathcal T={}&-\frac{2\mathcal G}{\beta^2t_2\epsilon^2}
{}_2F_1(-2\epsilon,1+2\epsilon;1-\epsilon;1-z^2),\nonumber\\
\mathcal N={}&\frac{4(1-z^2)\mathcal G}{\beta^2\epsilon(1-\epsilon)}
{}_2F_1(1+2\epsilon,1-2\epsilon;2-\epsilon;1-z^2),
\end{align}
where $\mathcal G=t_2^{-2\epsilon}\Gamma(1-\epsilon)\Gamma(1+\epsilon)\Gamma(1+2\epsilon)$. At $z=1$ both hypergeometric functions equal one, so $\mathcal T$ and $\mathcal N/(1-z^2)$ are finite at generic $\epsilon$. Here $-\boldsymbol q_2$ is parallel to $\boldsymbol u_1$, implying $\beta^2\mathcal W=2\omega_1y\mathcal V$. The $\mathcal N$ term in \eqref{eq:amplitude} therefore vanishes.

\vskip 4pt
With $\mathcal G$ factored out, the dimensionless combinations through order $\epsilon^0$ are
\begin{align}\label{eq:NT}
\frac{\beta^2\mathcal N}{\mathcal G}
&=-\frac{8\log z}{\epsilon}+8\log^2 z+4\Li(1-z^2),\nonumber\\
\frac{\beta^2t_2\mathcal T}{\mathcal G}
&=-\frac{2}{\epsilon^2}-\frac{8\log z}{\epsilon}+8\log^2 z+12\Li(1-z^2).
\end{align}
To obtain the finite Laurent coefficient of the amplitude, we expand $(4\pi)^{-D}\mu^{4\epsilon}\mathcal G$ through $\epsilon^2$. The scale appears as $(\mu^2/t_2)^{2\epsilon}$ in sector $22$ and $(\mu^2/t_1)^{2\epsilon}$ in sector $11$.

\vskip 4pt
A particular zero of the tree amplitude isolates the contribution of $\mathcal N$. At $t_1=t_2$, a polarization satisfying $yE_q+\mathcal V=0$ with $E_q\ne0$ makes both charge terms vanish even in the unexpanded tree amplitude \eqref{eq:exacttree}. The bracket in \eqref{eq:amplitude} then reduces to $\mathcal W\mathcal N/[\beta^2t_2(1-z^2)]$. The single-pole coefficient of $\mathcal N$ is $-8\log z/\beta^2>0$ for $0<z<1$. For the opposite charges and real polarization specified after \eqref{eq:exacttree}, the two residues have the same sign, so a simple pole survives in their sum.

\vskip 4pt
We independently check \eqref{eq:NT} through order $\epsilon^0$ by reducing the amplitude to master integrals using integration by parts~\cite{Chetyrkin:1981qh,Laporta:2000dsw} with reverse unitarity to preserve both cuts~\cite{Anastasiou:2002yz}. We solve the differential equations~\cite{Kotikov:1990kg,Remiddi:1997ny,Gehrmann:1999as,Henn:2013pwa} with boundary values from auxiliary mass flow~\cite{Liu:2017jxz} computed using AMFlow~\cite{Liu:2022chg,Huang:2026rjb}, and then test the solution at a second kinematic point (Appendix).

\vskip 6pt
\noindent{\bf Coulomb dressing.}~The two extra exchanges shift the transfer entering the radiative tree to $h_3=q_2-p$, where $p=h_1+h_2$. Write $\boldsymbol r,\boldsymbol s$ for the transverse components of $h_1,h_2$ and $\boldsymbol p=\boldsymbol r+\boldsymbol s$. The longitudinal component of $h_3$ contributes $\omega_1^2/\beta^2$ to its inverse propagator. The transverse projection $\boldsymbol q_\perp$ of $q_2$ has squared magnitude $t_2(1-z^2)$, giving the common integral
\begin{equation}\label{eq:J}
\mathcal J[f]=\int\!\frac{\dd^{d_\perp}\boldsymbol r\,\dd^{d_\perp}\boldsymbol s}{\pi^{d_\perp}}\,
\frac{f(\boldsymbol p)}
{\boldsymbol r^2\boldsymbol s^2[(\boldsymbol p-\boldsymbol q_\perp)^2+\omega_1^2/\beta^2]},
\end{equation}
with $\mathcal T=-2\mathcal J[1]/\beta^2$ and $\mathcal N=-2\mathcal J[\boldsymbol p\cdot\boldsymbol q_\perp]/\beta^2$.

\vskip 4pt
In the tree numerator, $q_2\mapsto q_2-p$ shifts $\mathcal W$ by $-2p^\mu F_{\mu\nu}u_1^\nu$ and leaves $\mathcal V$ fixed. Summing the cubic and contact diagrams yields the entire $Q_1^2Q_2$ term of \eqref{eq:tree} at the shifted transfer, while sector $11$ accounts for emission from the other line. Each extra exchange contributes the elastic kernel $K(\boldsymbol r)=-\ii Q_1Q_2y/(\beta\boldsymbol r^2)$, including the Jacobian $1/\beta$ from the two longitudinal constraints. With $k,u_1,u_2$ fixed and $\int_{\boldsymbol r}=\int\dd^{d_\perp}\boldsymbol r/(2\pi)^{d_\perp}$, we obtain
\begin{align}\label{eq:exponentiation}
\mathcal M^{(1)}_5&=\int_{\boldsymbol r}K(\boldsymbol r)\,\mathcal M^{(0)}_5(\boldsymbol q_\perp-\boldsymbol r),\nonumber\\
\mathcal M^{(2)}_5&=\frac12\int_{\boldsymbol r}\int_{\boldsymbol s}K(\boldsymbol r)K(\boldsymbol s)\,\mathcal M^{(0)}_5(\boldsymbol q_\perp-\boldsymbol r-\boldsymbol s),
\end{align}
where $q_1=k-q_2$ shifts oppositely and $1/2!$ accounts for the two identical exchanges. Both relations hold before expanding in $\epsilon$ (see Appendix for the one-loop derivation).

\vskip 4pt
A Fourier transform in the transverse transfer at fixed $k,u_1,u_2$ converts each convolution into a product. At impact parameter $b=|\boldsymbol b|>0$ the elastic kernel becomes
\begin{equation}\label{eq:phase}
\widetilde K(b)=2\ii\delta_0(b)
=-\frac{\ii Q_1Q_2y}{4\beta\pi^{1-\epsilon}}\,
 \Gamma(-\epsilon)(b^2)^\epsilon .
\end{equation}
Equivalently, $2\delta_0(b)$ is the transverse Fourier transform of the elastic tree amplitude in \eqref{eq:elastic} divided by $4m_1m_2\beta$. At leading long-wavelength order, the one- and two-loop amplitudes in impact-parameter space are the linear and quadratic terms in the expansion of $e^{2\ii\delta_0(b)}\widetilde{\mathcal M}_5^{(0)}$, respectively. The pole of this phase fixes the two-loop double pole. The finite parts of both polarization coefficients require the full convolution in \eqref{eq:exponentiation}, including exchange momenta comparable to the external transfers, where the tree numerator varies.

\vskip 6pt
\noindent{\bf Infrared structure.}~The infrared poles in \eqref{eq:J} arise when one or both of the additional exchange momenta become soft at fixed external transfer. When both exchanges become soft, the radiative tree is evaluated at the external transfer and $\mathcal T$ has a double pole. The numerator of $\mathcal N$ vanishes in this limit, eliminating its double pole. When only one exchange becomes soft, the other still shifts the tree numerator. For $\omega_1>0$ the third denominator is bounded below by $\omega_1^2/\beta^2$, and the integral is ultraviolet convergent near $D=4$.

\vskip 4pt
Scaleless self-interaction integrals can contain mutually canceling ultraviolet and infrared poles. For external scalars with unit pole residues, the Ward identity fixes the relative sign of the logarithmic vertex and external-line corrections. These cancel, including their two-loop overlap (Appendix). The use of pole masses and the Thomson definition of the charges fixes the remaining local terms at this order. All poles at $\epsilon=0$ in the leading amplitude at nonzero photon frequency are therefore infrared.

\vskip 4pt
The residue $b_{\mathrm{IR}}=\ii Q_1Q_2y/(4\pi\beta)$ of the pole in $2\ii\delta_0(b)$ is independent of $b$ and factors out of the inverse Fourier transform. With $\mu^{3\epsilon}$ stripped off, the tree amplitude \eqref{eq:tree} is independent of $\epsilon$ at fixed external kinematics and polarization contractions. Restoring $\mu^{2L\epsilon}$ and denoting the coefficient of $\epsilon^j$ by $\mathcal M^{(L)}_{5,j}$, we obtain
\begin{align}\label{eq:IR}
\mathcal M^{(1)}_{5,-1}&=b_{\mathrm{IR}}\mathcal M^{(0)}_5,\nonumber\\
\mathcal M^{(2)}_{5,-2}&=\tfrac12b_{\mathrm{IR}}^2\mathcal M^{(0)}_5,
&\mathcal M^{(2)}_{5,-1}&=b_{\mathrm{IR}}\mathcal M^{(1)}_{5,0}.
\end{align}
At a tree zero, the finite part of the one-loop amplitude thus fixes the two-loop single pole. Multiplying by the inverse pole part of the virtual Coulomb phase, $\exp(-b_{\rm IR}/\epsilon)$, removes the infrared poles from the leading long-wavelength amplitude through two loops~\cite{Weinberg:1965nx,Yennie:1961ad}.

\vskip 6pt
\noindent{\bf Soft-photon theorem.}~The infrared poles above come from virtual exchanges at fixed photon momentum. When the emitted photon itself becomes soft, the leading soft-photon theorem relates the five-point amplitude to elastic scattering~\cite{Low:1958sn,Weinberg:1965nx}. We take $k=\rho\hat k$ in a fixed direction, with $\rho\to0^+$ and $q_2\to\bar q\ne0$. Since $u_2\cdot q_2=0$ and $u_1\cdot q_2=u_1\cdot k\to0$, $\bar q$ is transverse to both velocities. Its magnitude remains small compared with both masses as $z\to0$. Expanding the incoming and outgoing soft factors to first order in the transfers gives
\begin{equation}\label{eq:softfactor}
S_\gamma=\sum_{a=1}^2\frac{Q_a}{m_a\omega_a}
\left(\frac{q_a\cdot\hat k}{u_a\cdot\hat k}\,\varepsilon\cdot u_a-q_a\cdot\varepsilon\right).
\end{equation}
The bracket remains finite; since $\omega_a=\rho\,u_a\cdot\hat k$, the soft factor has at most a simple pole in $\rho$. We compute the elastic amplitude independently at the same long-wavelength order by summing all orderings of three exchanged photons. Adding the two radiative charge sectors, we find
\begin{equation}\label{eq:softcheck}
\mathcal M^{(2)}_5
=S_\gamma\,\mathcal M^{(2)}_{4,\mathrm{el}}+o(\rho^{-1})
\end{equation}
with the elastic amplitude evaluated at $t_2=-\bar q^2$. We can take the soft limit in the convergence strip of \eqref{eq:allEpsilon} or after expanding about $\epsilon=0$. Both procedures yield the same soft residue order by order (Appendix).

\vskip 4pt
In sector $22$ the soft residue depends on $\mathcal N-t_2\mathcal T$. Permutation symmetry among the three exchanges in \eqref{eq:J} fixes its $z\to0^+$ limit. The explicit logarithms in \eqref{eq:NT} cancel in this combination. Subtracting its zero-frequency limit at fixed $y,t_2$ removes the remaining poles, leaving $8[\pi^2/6-\Li(1-z^2)]/\beta^2$ as $\epsilon\to0$. In four dimensions, logarithms of the photon energy can enter at subleading order without affecting the $1/\rho$ residue in \eqref{eq:softcheck}~\cite{Sahoo:2018lxl}.

\vskip 6pt
\noindent{\bf Conclusions and Discussion.}~The Coulomb dressing in \eqref{eq:exponentiation} accounts for the full leading long-wavelength term of the two-loop bremsstrahlung amplitude. At fixed external transfer the exchanges shift the tree numerator, thereby changing the relative polarization weights. We have evaluated both coefficients at nonzero photon frequency, including the finite terms left undetermined by the infrared poles.

\vskip 4pt
In gravity one can likewise sum over attachment orders on a massive line. If all orderings share a numerator and the net energy transfer vanishes in that line's rest frame, the eikonal denominators sum to delta functions. Explicit one-loop calculations of graviton emission~\cite{Georgoudis:2023lgf,Georgoudis:2023ozp} recover the iteration of elastic and radiative trees predicted by the gravitational eikonal description~\cite{DiVecchia:2022piu}. At two loops the challenge is to identify complete ordering sums in the presence of momentum-dependent tensor numerators and graviton self-interactions. In electrodynamics, a higher-loop calculation could test whether repeated elastic exchanges continue to exhaust the leading amplitude, as classical consistency suggests~\cite{Cristofoli:2021jas}.

\vskip 4pt
\noindent{\bf Acknowledgments.}~We thank Yu Jiao Zhu and Yongqun Xu for useful discussions. This work was supported in part by the Start-up Research Fund of Southeast University (No.\,RF1028624160).
YZ is supported by the National Natural Science Foundation of China (No.\,12575078 and 12247103). YW is supported by ANR-24-CE31-7996.
Large language models (primarily GPT-6 Astra and Kimi K3.1) were used under the authors' supervision to assist with diagram analysis and manuscript polishing.
The authors take full responsibility for the final manuscript and the results it reports.

\input{ref.bbl}

\clearpage
\appendix
\letterappendixtitle
\setlength{\parskip}{0pt}
\raggedbottom
\refstepcounter{section}
\label{app:details}

\begingroup
\setlength{\abovedisplayskip}{3pt}
\noindent\emph{Two-loop integral families.}~The three families in sector $22$ differ only in their linear denominators. For $f=1,2,3$, $\boldsymbol a\in\mathbb Z^9$ and $a_8,a_9\leq0$, define
\begin{align}\label{eq:family}
I^{(f)}_{\boldsymbol a}={}&\int\!\frac{\dd^D\ell_1\,\dd^D\ell_2}{(2\pi)^{2D}}\,
C_{a_1}(u_2\cdot\ell_1)\,C_{a_2}(u_2\cdot\ell_2)\nonumber\\
&\times\frac{(\ell_1\cdot\ell_2)^{-a_8}(\ell_1\cdot q_2)^{-a_9}}
{\prod_{j=3}^{7}\bigl(D_j^{(f)}\bigr)^{a_j}} .
\end{align}
For $n>0$, the cut factor is
\[
C_n(\xi)=\frac{(-1)^{n-1}}{(n-1)!}\,\delta^{(n-1)}(\xi).
\]
The integral vanishes if either cut index is nonpositive. An uncut propagator of index zero is omitted.

With $\ell_{12}=\ell_1+\ell_2$, the photon denominators are $D_3^{(f)}=-(q_2-\ell_{12})^2$, $D_4^{(f)}=-\ell_1^2$ and $D_5^{(f)}=-\ell_2^2$. The linear denominators are
\[
\begin{aligned}
\bigl(D_6^{(1)},D_7^{(1)}\bigr)&=\bigl(\omega_1-u_1\cdot\ell_1,\,\omega_1-u_1\cdot\ell_{12}\bigr),\\
\bigl(D_6^{(2)},D_7^{(2)}\bigr)&=\bigl(\omega_1-u_1\cdot\ell_{12},\,-u_1\cdot\ell_1\bigr),\\
\bigl(D_6^{(3)},D_7^{(3)}\bigr)&=\bigl(-u_1\cdot\ell_1,\,-u_1\cdot\ell_{12}\bigr).
\end{aligned}
\]
The amplitude is $\mathcal M_5^{(2)}|_{22}=\sum_{j=1}^9c_jI_j$, with
\[
\begin{aligned}
I_1&=I^{(2)}_{1,1,1,1,1,1,1,-1,0},\qquad
I_2=I^{(1)}_{1,1,1,1,1,1,1,-1,0},\\
I_3&=I^{(3)}_{1,1,1,1,1,1,1,-1,0},\qquad
I_4=I^{(2)}_{1,1,1,1,1,1,1,0,0},\\
I_5&=I^{(1)}_{1,1,1,1,1,1,1,0,0},\qquad
I_6=I^{(1)}_{1,1,1,1,1,1,0,0,0},\\
I_7&=I^{(3)}_{1,1,1,1,1,1,1,0,0},\qquad
I_8=I^{(1)}_{1,1,1,1,1,0,0,0,0},\\
I_9&=I^{(3)}_{1,1,1,0,1,1,0,0,0}.
\end{aligned}
\]
Including the cut Jacobians, the coefficients are
\begin{equation}\label{eq:coefficients}
\begin{aligned}[b]
&c_1=c_2=-2\widehat P,\quad c_3=\widehat P,\quad c_4=c_5=-\frac{K_2y^2U}{\omega_1H},\\
&c_6=\frac{2\omega_1\widehat P}{\beta^2},\quad c_7=\frac{K_2y^2R}{2\omega_1^2H},\\
&c_8=-\frac{(3D-11)\widehat P}{(D-4)\beta^2},\quad c_9=\frac{(2D-7)\widehat P}{2\omega_1(D-4)},
\end{aligned}
\end{equation}
where
\begin{equation*}
\begin{alignedat}{2}
&H=t_2\beta^2-\omega_1^2,
&\quad &K_2=8\pi^2Q_1^4Q_2^3m_2,\\
&P=\beta^2\mathcal W-2\omega_1y\mathcal V,
&\quad &\widehat P=K_2y^3P/(\omega_1^2H),\\
&U=\omega_1y\mathcal W-2(t_2+\omega_1^2)\mathcal V,
&\quad &R=t_2yP-2\omega_1U.
\end{alignedat}
\end{equation*}

The poles in $c_8,c_9$ require $I_8,I_9$ through first order in $\epsilon$. The cuts reduce $I_8$ to a massless sunset integral in one fewer dimension. Bubble integration in $I_9$ leaves a linear propagator with prescription $-\ii0$.

The two elliptic masters in each of families 1 and 3 have vanishing coefficients. An independent sum over cross-line exchange diagrams reproduces all nine coefficients in sector $22$. We obtain identical reductions in a common basis with Kira~\cite{Lange:2025fba} and with FIRE~\cite{Smirnov:2023yhb} using LiteRed symmetries~\cite{Lee:2013mka}. We verify the reduction and differential-equation matrices against exact IBP identities. Boundary data at $(\beta,t_2/\mu^2,\omega_1/\mu)=(1,4,1)$ fix the solution. At $(2,9,9/2)$, we check 17 nonelliptic master-integral entries across the three families and both moments through $\epsilon^0$ against independent AMFlow evaluations at 216-digit working precision. Of 90 absolute differences from reference values $v_{\rm ref}$, each normalized by $\max(1,|v_{\rm ref}|)$, the largest is $3.1\times10^{-35}$.

\endgroup

\vskip 4pt
\noindent\emph{Ordering sum on the emitting line.}~The line-2 cuts and photon propagators are symmetric in the three exchanges. Terms containing scalar products of distinct exchanged momenta have support only at zero frequency after summing over the four cubic emission positions. The ordered sum for a seagull joining two virtual photons has the same support. At positive frequency, we order emission relative to the $h_3$ exchange and sum over the $h_1,h_2$ attachments, leaving their times unrestricted. Together with \eqref{eq:eikonal}, Fourier transformation yields $(2\pi\ii)^2\delta(u_1\cdot h_1)\delta(u_1\cdot h_2)$. The six orderings with emission and $h_3$ at a common seagull yield the same factor.

For independent real $a,b$, the denominators $a$ and $b-a$ have the same negative imaginary regulator, so their sum has twice that regulator. Taking the real part as the regulator vanishes in the sense of distributions in $(a,b)$ gives
\begin{align}\label{eq:PB}
\PV\frac1a\PV & \frac1{b-a}-\PV\frac1b\PV\frac1{b-a}\nonumber\\
&=\PV\frac1a\PV\frac1b-\pi^2\delta(a)\delta(b).
\end{align}

Relabeling the photons eliminates the principal-value products, leaving the delta functions in \eqref{eq:PB}.

By rotational symmetry, the vector first moment is parallel to $q_\perp=q_2-(\omega_1/\beta^2)(yu_2-u_1)$. For nonzero transverse transfer, its coefficient is the moment with numerator $\boldsymbol p\cdot\boldsymbol q_\perp$ divided by $\boldsymbol q_\perp^2$. Using $2q_\perp^\mu F_{\mu\nu}u_1^\nu=P/\beta^2$ and the measure conversion $\pi^{D-2}/(2\pi)^{2(D-2)}$, we obtain \eqref{eq:amplitude} from \eqref{eq:exponentiation}.

\vskip 4pt
\noindent\emph{Evaluation in $D$ dimensions.}~At fixed $\boldsymbol p$, the $\boldsymbol r$ integral in \eqref{eq:J} gives the massless bubble
\[
\frac{\Gamma(1+\epsilon)\Gamma(-\epsilon)^2}
     {\Gamma(-2\epsilon)(\boldsymbol p^2)^{1+\epsilon}}.
\]
Writing $a=\omega_1^2/\beta^2$, we introduce a Feynman parameter and shift $\boldsymbol p$ to obtain
\begin{align}\label{eq:parameter}
\mathcal T={}&-\frac{2\Gamma(-\epsilon)^2\Gamma(1+2\epsilon)}{\beta^2\Gamma(-2\epsilon)}
\int_0^1\!\frac{x^\epsilon(1-x)^{-1-2\epsilon}\,\dd x}
{[a+x(t_2-a)]^{1+2\epsilon}}.
\end{align}
Only the even part $(1-x)(t_2-a)$ of the shifted first-moment numerator survives integration. The substitution $x\mapsto1-x$ yields Euler's hypergeometric form. The original factor $1-x$ raises one beta-function argument by one, producing the prefactor ratio $-2\epsilon/(1-\epsilon)$ in \eqref{eq:allEpsilon}.

\vskip 4pt
\noindent\emph{Tree amplitude.}~Summing the six tree diagrams before the long-wavelength expansion, we obtain
\begin{align}\label{eq:exacttree}
\mathcal M^{(0)}_{5,\mathrm{exact}}={}&\frac{2Q_1^2Q_2M_2}{t_2}
\frac{M_1^2(y\mathcal W-2\omega_1\mathcal V)+\Delta\mathcal W_2/8}
{M_1^2\omega_1^2-\Delta^2/16}\nonumber\\
&+(1\leftrightarrow2),
\end{align}
where $\mathcal W_2$ denotes $\mathcal W$ with the particle labels exchanged. The leading term is \eqref{eq:tree}. For $u_2=(1,0,0,0)$, $u_1=(5/4,0,0,3/4)$, $k=(3\kappa/5)(1,1,0,0)$ and $q_2=\kappa(0,0,1,-1)$, with $0<\kappa\ll m_a$, the transfers satisfy $t_1=t_2=2\kappa^2$. The unnormalized polarization $\varepsilon=(0,0,16/25,1)$ then yields $yE_q+\mathcal V=0$. For equal masses $m$ and charges $(Q,-Q)$, the two-loop single-pole residues in $22$ and $11$, in units of $Q^7m/(\pi^2\kappa^2)$, are $(20/27)\log 2$ and $(625/918)\log(25/8)$, respectively.
At $t_1=t_2$, $\mathcal W/\omega_1=-\mathcal W_2/\omega_2=-2E_q$. The charge factors also have opposite signs, so the residues add.

\vskip 4pt
\noindent\emph{One-loop Coulomb dressing.}~Summing both orientations in \eqref{eq:eikonal} cuts the line-1 propagator. Together, the two matter cuts leave the measure $\ii\int_{\boldsymbol r}/(2\pi\beta)$. With $D_{\boldsymbol r}=(\boldsymbol r-\boldsymbol q_\perp)^2+\omega_1^2/\beta^2$, $B=\int_{\boldsymbol r}1/(\boldsymbol r^2D_{\boldsymbol r})$ and $J=\int_{\boldsymbol r}1/D_{\boldsymbol r}$, the amplitude is
\begin{equation}\label{eq:lower}
\left.\mathcal M^{(1)}_5\right|_{Q_1^3Q_2^2}
=-\frac{\ii Q_1^3Q_2^2m_2}{\beta\omega_1^2H}\left(yR B-y^2P J\right).
\end{equation}
Since $2\boldsymbol r\cdot\boldsymbol q_\perp=\boldsymbol r^2+t_2-D_{\boldsymbol r}$, twice the first moment is $t_2B+\int_{\boldsymbol r}(1/D_{\boldsymbol r}-1/\boldsymbol r^2)$. The subtraction removes the ultraviolet tail. At small momentum, $D_{\boldsymbol r}$ tends to $t_2$, so the bubble cancels the subtracted massless propagator. The first moment remains finite at $D=4$. Only $B$, multiplying the unshifted tree numerator, carries the infrared pole in \eqref{eq:IR}.
The scaleless integral of $1/\boldsymbol r^2$ vanishes in dimensional regularization, yielding \eqref{eq:lower}. Particle exchange gives the other sector.
For $\mathcal G_1=(4\pi/t_2)^\epsilon\Gamma(1+\epsilon)$, we find, through order $\epsilon^0$, $4\pi t_2 B/\mathcal G_1=-1/\epsilon-2\log z$ and $4\pi J/\mathcal G_1=1/\epsilon-2\log z$.

Expanding $\mu^{2\epsilon}\mathcal G_1$ and $(4\pi)^{-D}\mu^{4\epsilon}\mathcal G$ reproduces \eqref{eq:IR}, including its single-pole $\log z$ terms.

\vskip 4pt
\noindent\emph{Elastic amplitude and soft limit.}~We independently sum all photon orderings on both massive lines through two loops. Seagulls are suppressed at this order, and the transverse integrals give
\begin{align}\label{eq:elastic}
\mathcal M^{(L)}_{4,\mathrm{el}}={}&-\frac{4m_1m_2(-\ii)^L(Q_1Q_2y)^{L+1}}{(L+1)!\,\beta^L}\,B_L,\nonumber\\
B_L={}&\frac{t_2^{-1-L\epsilon}}{(4\pi)^{L(1-\epsilon)}}\,
\frac{\Gamma(1+L\epsilon)\Gamma(-\epsilon)^{L+1}}{\Gamma(-(L+1)\epsilon)},
\end{align}
with $B_0=1/t_2$. The one-loop pole has residue $b_{\rm IR}$ times the elastic tree amplitude.
At zero frequency, the three exchange momenta in \eqref{eq:J} enter symmetrically and sum to the transverse transfer. Accordingly, $\mathcal N$ equals two thirds of $t_2\mathcal T$ in the convergence strip. With $\mathcal J[1]=(4\pi)^{2-2\epsilon}B_2$, continuation in $\epsilon$ gives
\begin{equation}\label{eq:softNT}
(\mathcal N-t_2\mathcal T)\big|_{z=0}=\frac{32\pi^2t_2}{3\beta^2}\,(4\pi)^{-2\epsilon}B_2.
\end{equation}
For $\mathcal N-t_2\mathcal T$, the numerator cancels one power of the bracket in \eqref{eq:parameter} with a minus sign. Euler's integral yields $[2\mathcal G/(\beta^2\epsilon^2)]\,{}_2F_1(-2\epsilon,2\epsilon;1-\epsilon;1-z^2)$. The first nonanalytic correction scales as $z^{2-2\epsilon}$ around zero, so expanding about $\epsilon=0$ preserves the residue in \eqref{eq:softcheck}.

\vskip 4pt
\noindent\emph{Self-interactions and hard regions.}~The logarithmic self-interaction integral on line $a$ is
\begin{equation}\label{eq:self}
\begin{aligned}[b]
&\mathcal S_a=\int\!\frac{\dd^D\ell}{(2\pi)^D}\frac1{(-\ell^2)(u_a\cdot\ell)^2},\\
&\mathcal S_a\big|_{\mathrm{poles}}=\frac{\ii}{8\pi^2}\left(\frac1{\epsilon_{\mathrm{UV}}}-\frac1{\epsilon_{\mathrm{IR}}}\right).
\end{aligned}
\end{equation}
At zero transfer, the logarithmic vertex correction is $V_a=-\ii Q_a^2\mathcal S_a$. Including seagulls, one such self-interaction on each line multiplies the tree by $V_1V_2$. With unit external-line residue, the Ward identity yields $\delta Z_a=-V_a$, canceling the overlap $(V_1+\delta Z_1)(V_2+\delta Z_2)$, including its mixed UV--IR pole. The pole-mass condition fixes the local mass term. Defining the charge in the Thomson limit subtracts the vacuum polarization at zero momentum.

In all 32 diagrams with two photons joining the same seagull pair, the three photon propagators scale as $\lambda^{-6}$. The degree of the soft integrand is at least $-9$ for the 24 diagrams with three internal scalar propagators and $-8$ or higher for the eight with two.

When a hard subgraph joins both lines and one loop remains soft, one photon and at most three matter propagators produce an integrated degree of $4-2-3=-1$ or higher in four dimensions. A fully hard scattering subgraph acquires at most one soft pole from external emission. Both are subleading to degree $-2$.

In \eqref{eq:J} at $D=4$, the ultraviolet degrees are $-2$ for the scalar integral and at most $-1$ for its first moment when either momentum grows large or both scale together in generic directions. At a fixed sum, the numerator remains unchanged, resulting in degrees at most $-2$ for both. Thus \eqref{eq:J} is ultraviolet finite near $D=4$. At positive frequency, its poles at $\epsilon=0$ are infrared.

\end{document}

%% file: ref.bbl
\providecommand{\href}[2]{#2}\begingroup\raggedright\endgroup